\documentclass[manuscript,screen]{acmart}
\AtBeginDocument{%
  \providecommand\BibTeX{{%
    \normalfont B\kern-0.5em{\scshape i\kern-0.25em b}\kern-0.8em\TeX}}}

\usepackage{caption} 
\usepackage{subcaption}
\usepackage{paralist}
\newcommand{\blu}[1]{\textcolor{black}{#1}}
\setcopyright{none}
\renewcommand\footnotetextcopyrightpermission[1]{}
\begin{document}

\title{Marching with the Pink Parade:  Evaluating Visual \\ Search  Recommendations for Non-binary Clothing Items}

\author{Siddharth D Jaiswal}
\email{siddsjaiswal@kgpian.iitkgp.ac.in}
\author{Animesh Mukherjee}
\email{animeshm@iitkgp.ac.in}
\affiliation{
	\institution{Dept. of Computer Science and Engineering, Indian Institute of Technology}
	\city{Kharagpur}
	\state{West Bengal}
	\country{India}
}

\renewcommand{\shortauthors}{Jaiswal and Mukherjee}
\begin{abstract}
    Fashion, a highly subjective topic is interpreted differently by all individuals. E-commerce platforms, despite these diverse requirements, tend to cater to the average buyer instead of focusing on edge cases like non-binary shoppers. This case study\footnote{See video demo here: https://tinyurl.com/ddv886h2}, through participant surveys, shows that visual search on e-commerce platforms like Amazon, Beagle.Vision and Lykdat, is particularly poor for non-binary clothing items. Our comprehensive quantitative analysis shows that these platforms are more robust to binary clothing inputs. The non-binary clothing items are recommended in a haphazard manner, as observed through negative correlation coefficients of the ranking order. The participants also rate the non-binary recommendations lower than the binary ones. Another intriguing observation is that male raters are more inclined to make binary judgements compared to female raters. Thus it is clear that these systems are not inclusive to the minority, disadvantaged communities of society, like LGBTQ+ people. We conclude with a call to action for the e-commerce platforms to take cognizance of our results and be more inclusive.\footnote{\textcolor{red}{This work has been accepted for publication at ACM CHI 2022 (Case Studies) as an Extended Abstract}}
\end{abstract}



\maketitle
	\section{Introduction}
	Fashion permeates every aspect of modern human life, ranging from clothing and lifestyle to culture and societal behaviour. People shop for clothes, apparel, and footwear from both brick and mortar stores and online e-commerce platforms. The fashion industry, projected to be worth \$3.3 trillion by 2030~\cite{ecomm_rev}, is a substantial part of the global economy and affects many livelihoods in the production and supply chains.
		
	Although there has been a lot of stigma against LGBTQ+ people historically and almost no concern for their rights in any aspect of societal life, in recent time there is a growing understanding and acceptance of gender fluidity and homosexuality~\cite{article_377}. 
	
	As with other commodities and services (for e.g., access to public washrooms, changing rooms in clothing stores, etc), clothing items have been segregated on the basis of binary genders -- male and female with no recognition for trans or other gender fluid needs or preferences. LGBTQ+ people require safe spaces in shared public areas to deter harassment and discrimination~\cite{weinhardt2017transgender}, which such establishments may not be able to provide. Thus, a person from a disadvantaged community might prefer shopping for clothes and other items online from the safety and comfort of their homes.
	
	A huge gap exists in this space as none of the major e-commerce platforms offer a non-binary option when shopping for clothing items or apparel~\cite{amazon_home,flipkart_home}, which means that an LGBTQ+ person has no option but to search for clothes in the Men's section or the Women's section. Also, there is no way for the said person to verify the fit or size of the clothing item as the clothes are worn by cis-gendered models and aren't representative of their needs. Thus, we observe a clear need of bridging this gap by the e-commerce platforms such that they can provide a holistic shopping experience to LGBTQ+ members of society by not just selling clothing items and apparel relevant to them but also improve their shopping experience.
	
	\subsection{Visual search}
	Various online platforms and services support a special form of search called \textit{visual search} which allows a user to upload an image and get recommendations relevant to the uploaded image. This can be used for searching for normal items like, Google Lens~\cite{google_lens} and Pinterest Lens~\cite{pinterest_lens} or specifically for clothing items, like Amazon's StyleSnap~\cite{amz_stylesnap}. Thus visual search provides an extra dimension for searching items for users who might not be able to provide the exact textual phrase or if text-based search is not available in their preferred language. Similarly, visual search may show more relevant results as the ML/DL model running in the backend has access to more visual features it can use as compared to textual features. Unfortunately, none of the models in use by commercial platforms are available publicly and there is no definitive way of determining how the search is working for any given service. Thus, there is a potential for various forms of bias~\cite{mehrabi2021survey} like evaluation bias, population bias, behavioural bias, user interaction bias and emergent bias amongst others. 
	
	In this study, we explore possible \textit{user interaction} and \textit{emergent} biases in visual search for various clothing items, including and specially gender-neutral ones, by using expert and normal user opinions to identify whether the search results are relevant to the input images on a Likert-like scale and also what the rater's perception is about the gender relevance of the recommended clothing item (either male/female or non-binary). We perform this study on three e-commerce platforms that allow visual search -- Amazon StyleSnap~\cite{amz_stylesnap}, Beagle.Vision~\cite{bgl_vision} and Lykdat~\cite{lykdat}.
	 
	\subsection{Research questions} In this section, we state the research questions we will attempt to address in this case study.
	
	\noindent{\textbf{RQ1}} Whether the raters' consider the recommendations for the different types of input clothing items to be binary or non-binary? Through this question, we our intent is to identify how users' perceive the gender of clothing and whether the general opinion is inclined towards a non-binary or a binary categorization.
	
	\noindent{\textbf{RQ2}} How relevant the recommendations are for any given input clothing item? Through this question, we attempt to understand how well the visual search recommendation system works for each of the three platforms identified above, using a Likert-like scale. We use the recommendation scores to identify the relevance across different types of gendered clothing items, and also to rank the three platforms themselves in order of best to worst. We further check whether the recommendation scores of the items are correlated to their rank in the search results.
	
	\noindent{\textbf{RQ3}} How are the male raters' responses in comparison to the female raters' responses? \blu{Specifically, we study how the males respond to questions pertaining to both gender and relevance for the given clothing items vs. how the females respond to them and attempt to understand their opinions about the results of the visual search on the three platforms.}
	\section{Literature Survey}
We discuss some of the relevant prior literature studying fashion from a social computing perspective as well some of the existing work in visual search in recommendation systems. 

\subsection{Fashion in computing}
Design and style being subjective topics, make it inherently difficult to design systems that can define what is fashionable. Therefore, most research involves human annotators. A recent work by Ragot et al.~\cite{ragot_chi20} used this to compare art generated by humans and AI. The social computing community has also made attempts to understand what type of designs or fashion appeal to humans. Park et al.~\cite{park_cscw16} evaluated the success of models in the fashion industry based on their presence on Instagram. Similarly, Kiapour et al.~\cite{kiapour_eccv14} were able to link personal style with clothing behaviour. Yamaguchi et al.~\cite{yamaguchi_mm14} evaluated visual behaviour on fashion social networks and Maity et al.~\cite{maity_icwsm19} analyses the content in images on Pinterest fashion boards to identify the driving factors for their popularity. 
\blu{ While none of these works focus on non-binary clothing being sold in the market, our focus is on precisely this because we believe that the existing system is far too broken. This was seen in a recent related study by Dash et al.~\cite{dash2021umpire} who identified a significant amount of exposure bias in the recommendations by Amazon. Our study takes inspiration from this work and we feel that a careful study should be performed on such systems and these should be fixed before new technology can be installed.}
For example, Devendorf et al.~\cite{devendorf_chi16}, in their CHI'16 work, explored how dynamic textile displays are affected by personal style choices while a Pan et al.~\cite{pan_dis12} attempt to reconceptualize fashion in the domain of sustainable HCI. More recently, works~\cite{sun_chi21,tepe_chi21} have appeared in CHI'20 exploring various methods for fashion design using novel methods.

In this case study, we identify that the industry and users as a whole need to take a step back and evaluate the existing biases in genderfluid fashion choices before designing for the future. Thus our work sits as a parallel to the existing work on fashion in computing.

\subsection{Visual Search}
Visual search is found on search~\cite{google_lens}, image-sharing~\cite{pinterest_lens} and social media~\cite{snapchat_scan} applications. Lately, retailers have been using it to enhance the shopping experience of customers on e-commerce platforms~\cite{shopify_vissearch,curatti_vissearch,medium_vissearch}. Amazon's Stylesnap~\cite{amz_stylesnap} feature, launched in 2019, allows users to upload images of clothing items and receive relevant recommendations for the same. Beagle.Vision~\cite{bgl_vision} and Lykdat~\cite{lykdat} work in a similar way but none of these platforms have released their model architectures or training datasets. This impedes a better understanding of their algorithms and raises concerns of privacy and fairness~\cite{privacy_vissearch}.

\blu{In this work, we attempt to investigate such fairness concerns. To this end, we use images of individuals wearing different types of clothing items. These individuals may belong to different gender groups.}

    	\section{Dataset and Methodology}
	In this section we first discuss the details of the dataset used for the study, followed by a description of the platforms evaluated as part of the study. We next describe the demographics of the participants and finally the format of collecting and evaluating user responses.
	
	\subsection{Dataset}
	
		\begin{table}[t]
		\begin{minipage}{0.48\textwidth}
		\noindent
		\footnotesize
		\centering
		\begin{tabular}{ |c|l|}
			\hline
			{\bf Gender Category} & {\bf Type of Clothing Item}\\
			\hline
			Male-centric & bottomwear | shirt | T-shirt | suit \\
			\hline
			Female-centric & bottomwear | shirt | top | dress | formalwear \\
			\hline
			Non-binary & casualwear | dress | dungarees | suit | formalwear\\
			\hline
		\end{tabular}	
		\caption{\footnotesize{\bf Type of clothing items collected per gender category in the dataset. There are 8 unique items for each type leading to a total of 112 items.}}
		\label{Tab: dataset-clothing}
		\vspace{-5 mm}
		\end{minipage}
		\begin{minipage}{0.48\textwidth}
		\noindent
		\footnotesize
		\centering
		\begin{tabular}{ |c|c|c|c|}
			\hline
			{\bf Age} & {\bf M} & {\bf F} & {\bf Professional Background}\\
			\hline
			19-25 & 5 & 6 & Undergraduate College Student \\
			\hline
			26-30 & 1 & 2 & Research Scholar \& Working Professional\\
			\hline
			31-35 & 1 & 0 & Working Professional \\
			\hline
		\end{tabular}	
		\caption{\footnotesize {\bf Demographics of the participants who took part in the survey. The majority were undergraduate college students and females.}}
		\label{Tab: user-demographic}
		\vspace{-5 mm}
		\end{minipage}
	\end{table}
	
	\begin{figure}[t]
		\centering
		\begin{subfigure}{0.32\columnwidth}
			\centering
			\includegraphics[height=4cm]{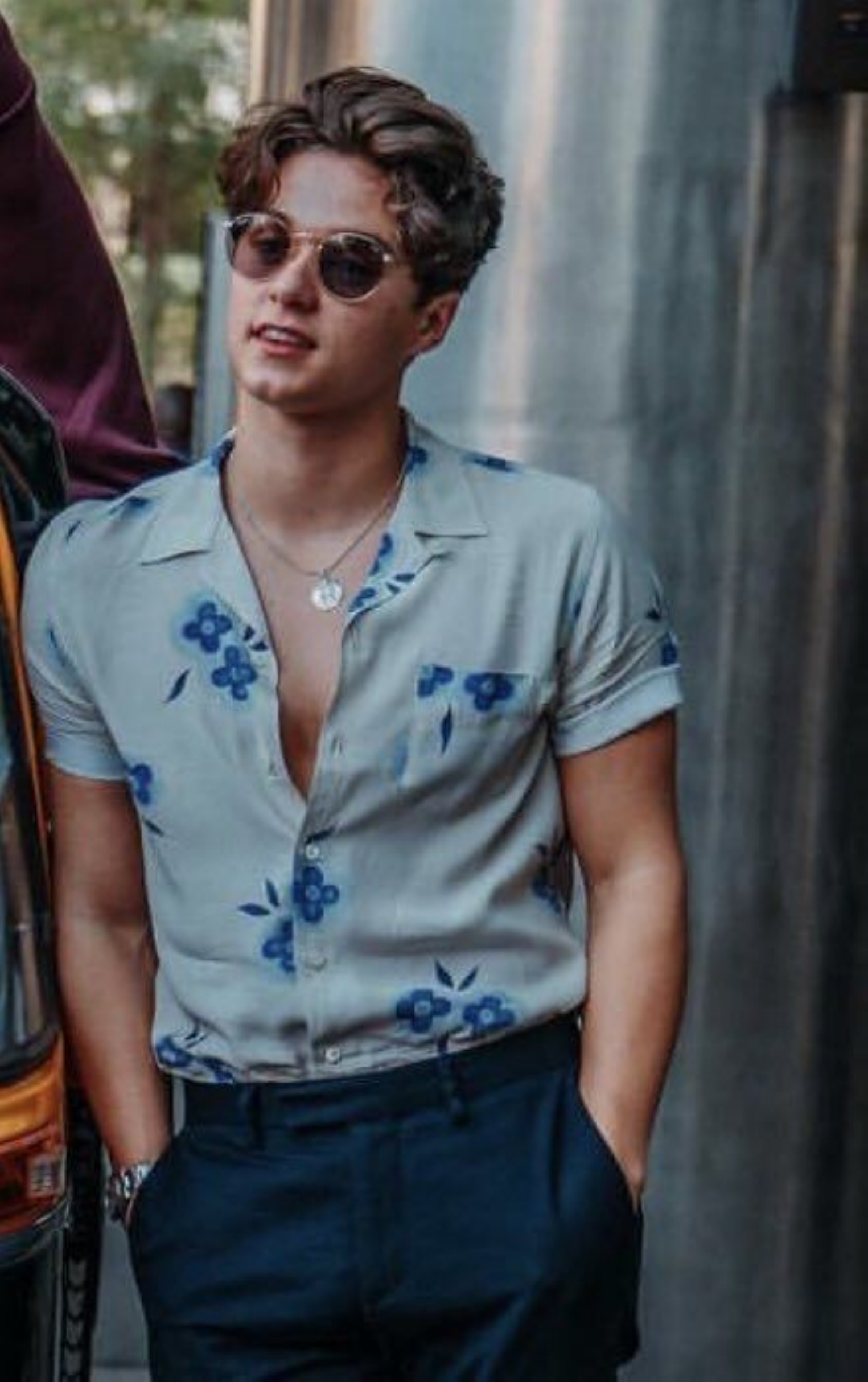}
			\caption{}
			\label{img1:a}
		\end{subfigure}%
		~\begin{subfigure}{0.32\columnwidth}
			\centering
			\includegraphics[height=4cm]{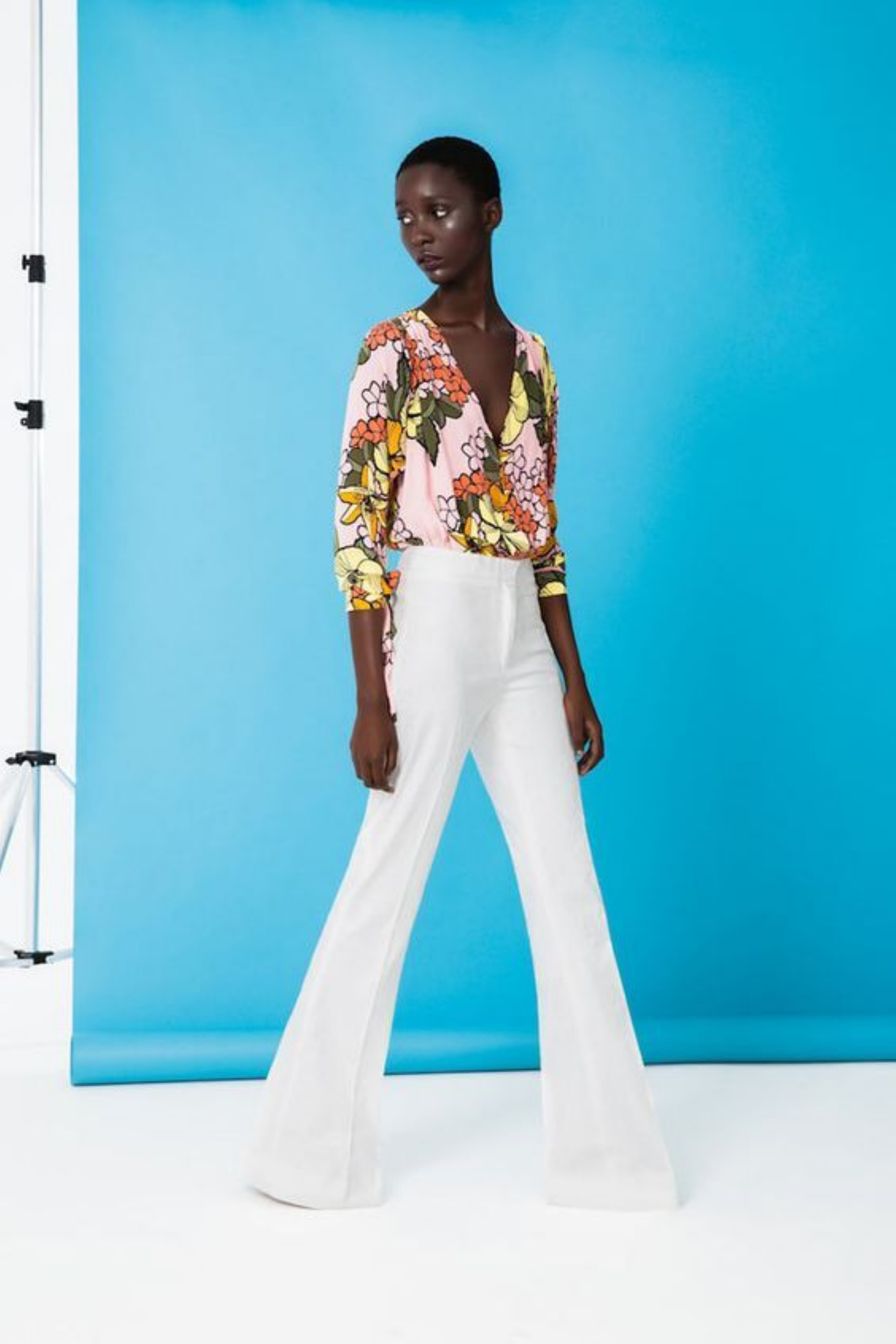}
			\caption{}
			\label{img1:b}
		\end{subfigure}
		~\begin{subfigure}{0.32\columnwidth}
			\centering
			\includegraphics[height=4cm]{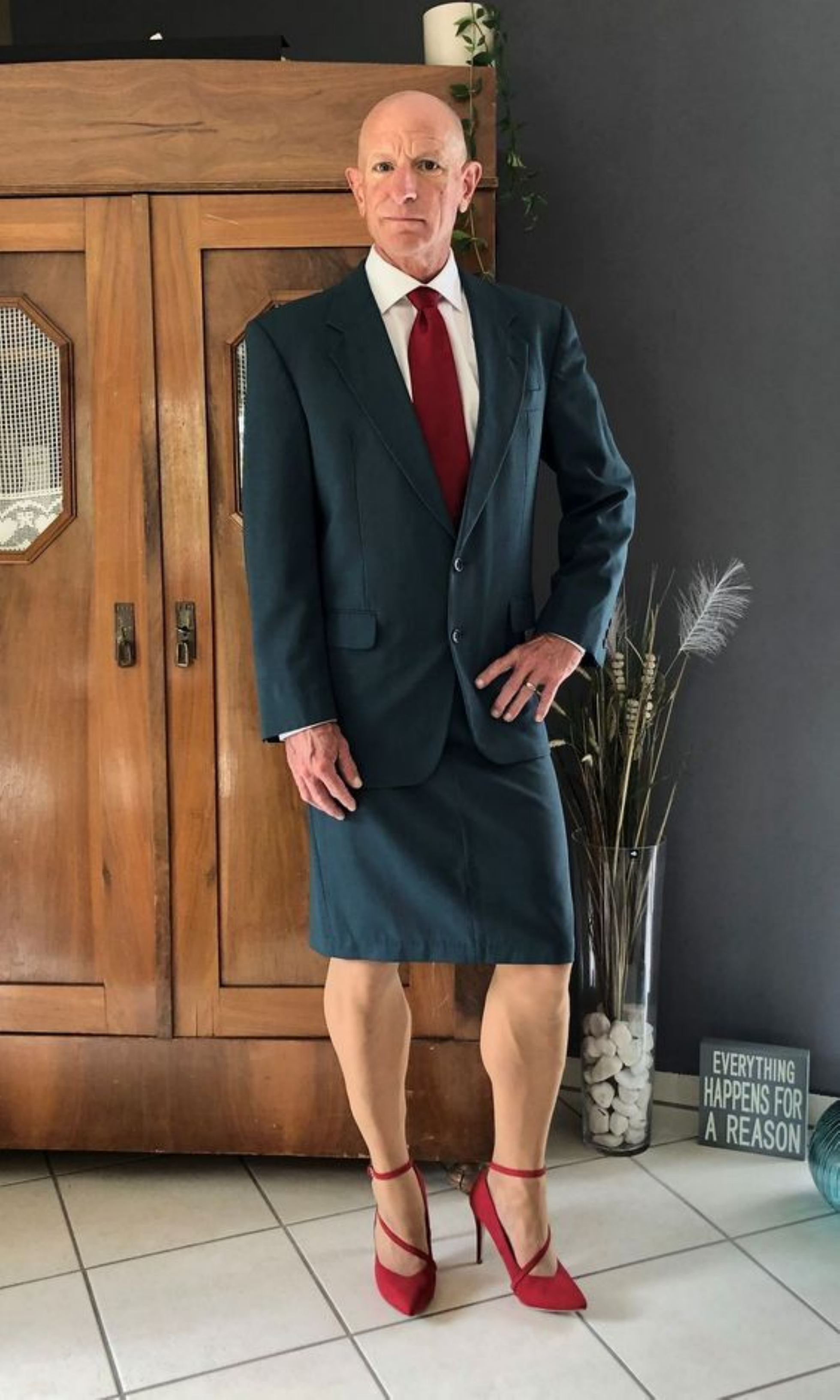}
			\caption{}
			\label{img1:c}
		\end{subfigure}
		
		\caption{\footnotesize{\bf Images collected from Pinterest website. (a) is an example of a male-centric shirt, (b) is an example of a female-centric bottomwear, (c) is an example of a non-binary suit.}}
		\label{img1: clothing-example}
		\vspace{-5 mm}
	\end{figure}
	
	We collected 112 images of various forms of clothing items from the Pinterest~\cite{pinterest} website by searching for specific queries like `\textit{mens jeans}', `\textit{womens tops}' and `\textit{androgynous clothing}' between July and September, 2021. 
	\blu{We then distributed the collected clothing items into three categories through a manual inspection of the data followed by a detailed discussion between the authors and adjudication by a fashion design expert. The resolved categories were \textit{male-centric} clothing, \textit{female-centric} clothing and \textit{non-binary} clothing.}
	Clothes in each gender category were further distributed into buckets corresponding to different clothing types -- `bottomwear', `T-shirts', `suits', etc. More details regarding the categories and count of images per category are noted in Table~\ref{Tab: dataset-clothing} and images of some of the clothing items collected from Pinterest are shown in Figure~\ref{img1: clothing-example}.

	\subsection{Platforms}
	
	\begin{figure}[t]
		\centering
		\begin{subfigure}{0.48\columnwidth}
			\centering
			\includegraphics[height=4.5cm]{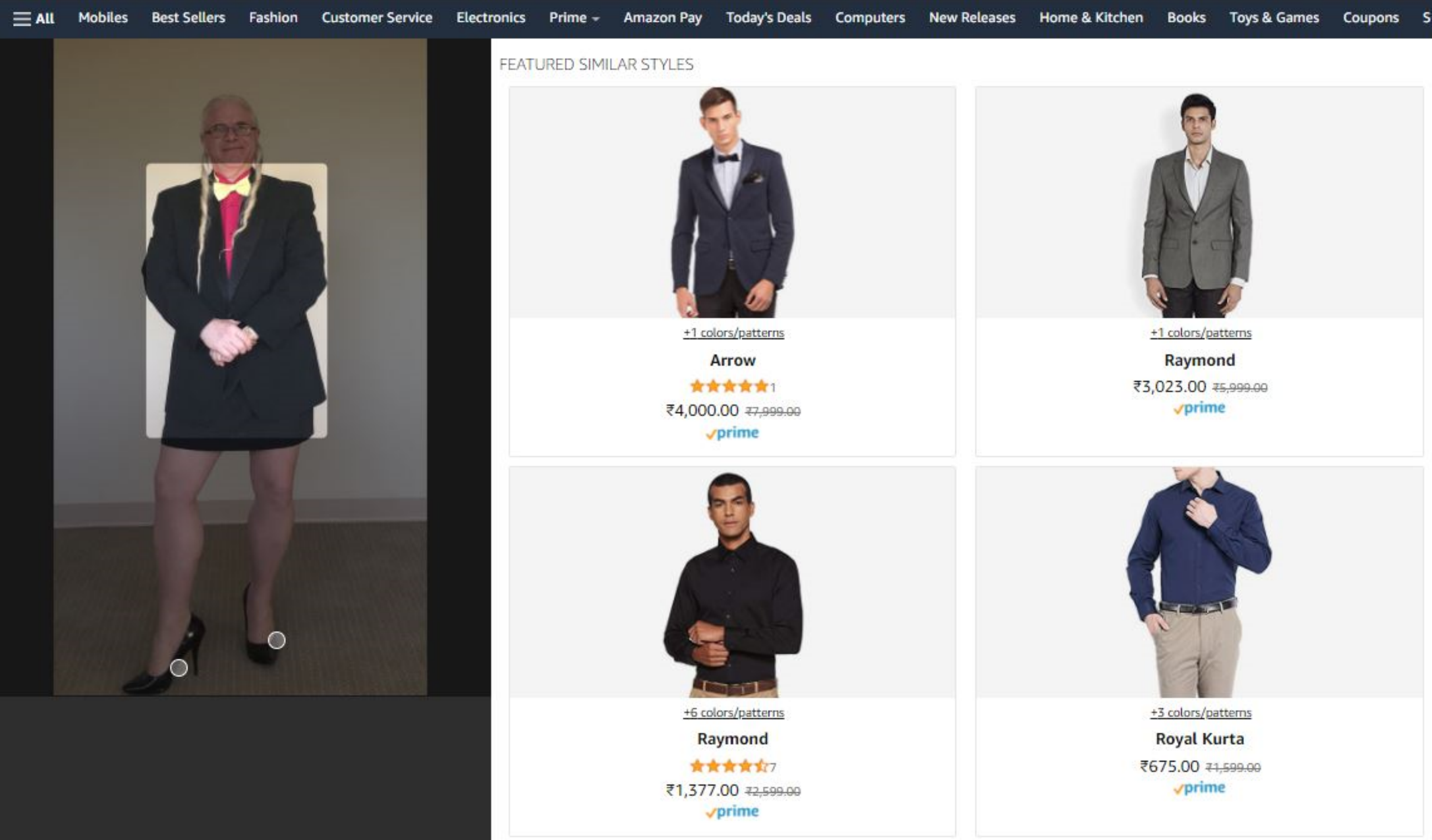}
			\caption{}
			\label{img2:a}
		\end{subfigure}%
		\begin{subfigure}{0.48\columnwidth}
			\centering
			\includegraphics[height=4.5cm]{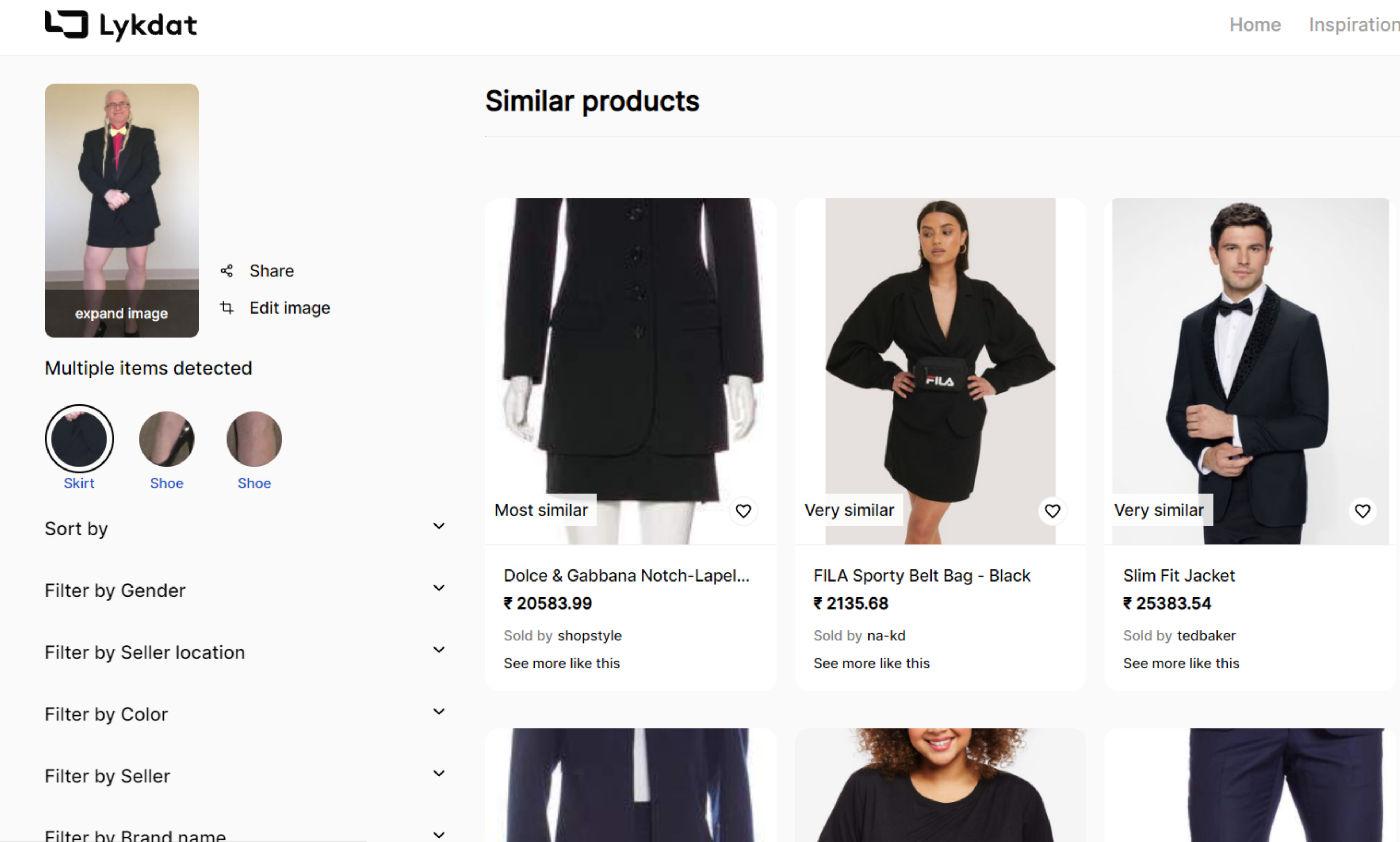}
			\caption{}
			\label{img2:b}
		\end{subfigure}%
	\\
		~\begin{subfigure}{0.48\columnwidth}
			\centering
			\includegraphics[height=4.5cm]{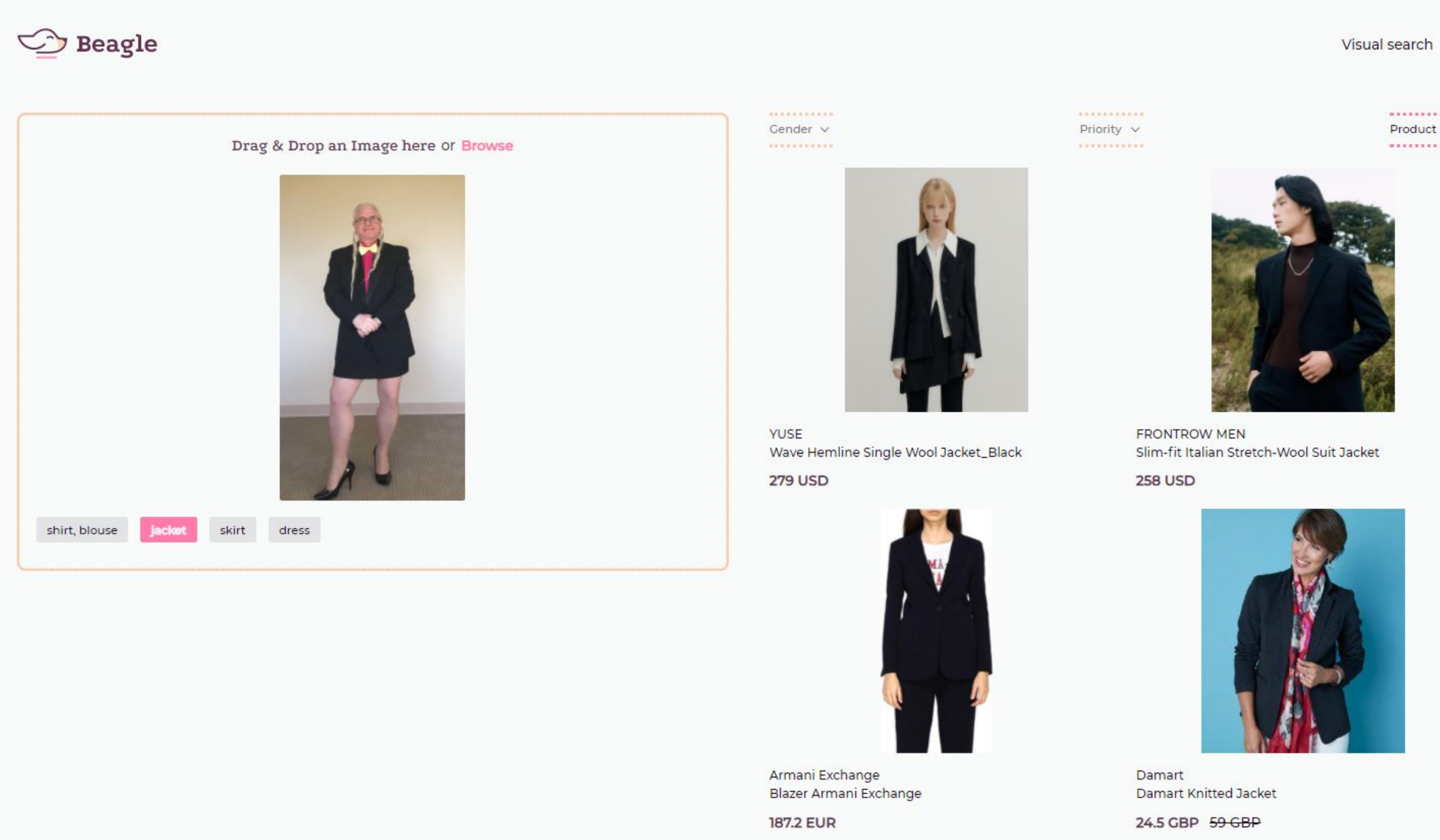}
			\caption{}
			\label{img2:c}
		\end{subfigure}
		~\begin{subfigure}{0.48\columnwidth}
			\centering
			\includegraphics[height=5.5cm]{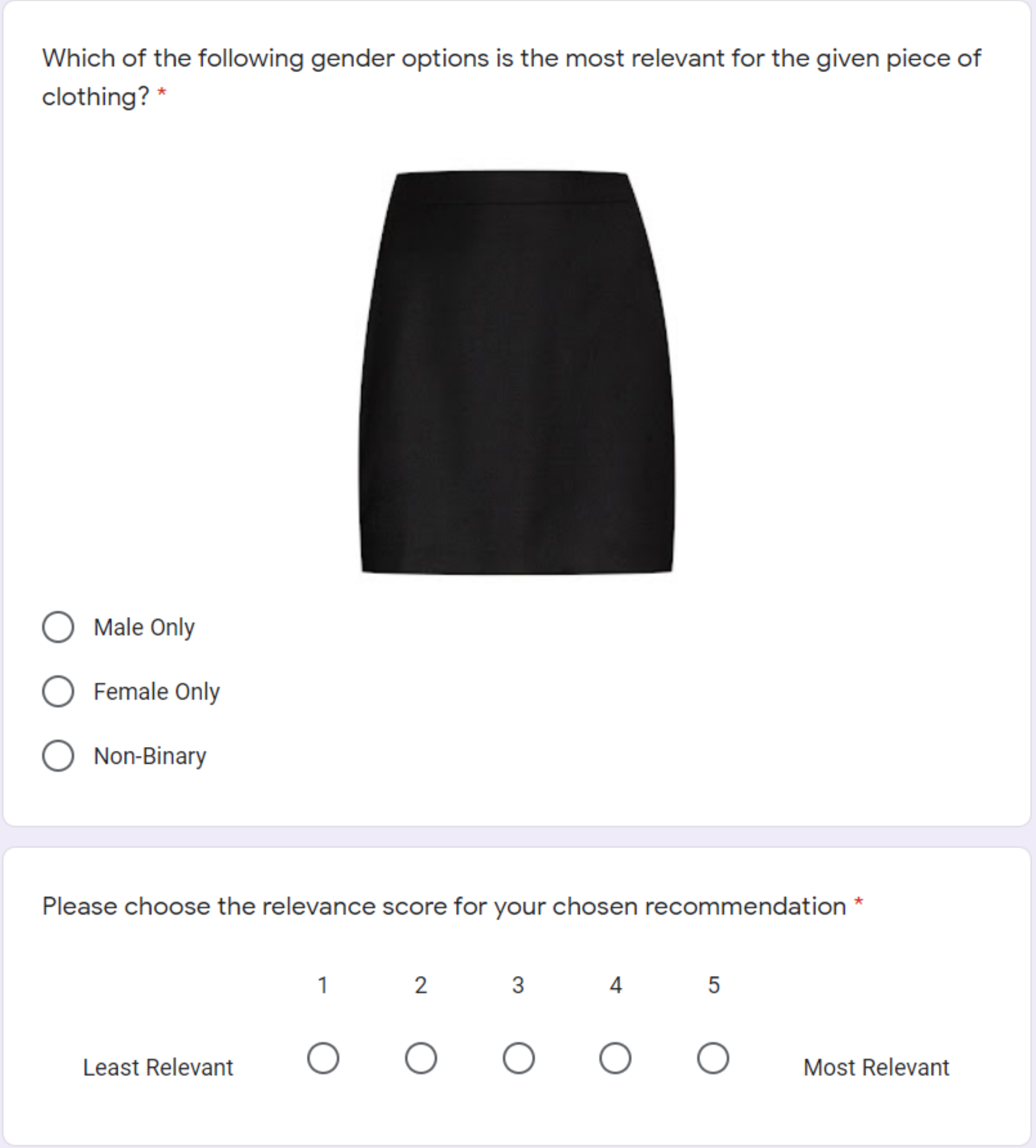}
			\caption{}
			\label{img2:d}
		\end{subfigure}
		
		\caption{\footnotesize {\bf Search results from the three platforms and a snapshot of the Google form shared with the participants. (a) is for Amazon Stylesnap, (b) is for Lykdat, (c) is for Beagle.Vision. Amazon recommends male centric clothing with shirts and suits mixed together whereas both Beagle and Lykdat recommend both male and female centric clothing. (d) is a snapshot of one of the Google forms with the two questions for every \textit{output image}.}}
		\label{img2: search-results}
		\vspace{-5 mm}
	\end{figure}

	\blu{Three platforms were evaluated for their visual search recommendations as part of this case study -- Amazon \textsc{StyleSnap}~\cite{amz_stylesnap}, because of its pervasiveness in the consumer industry, and \textsc{Beagle.Vision}~\cite{bgl_vision} and \textsc{Lykdat}~\cite{lykdat} because, despite being relatively lesser known companies, their visual search tech is openly accessible and is being used by customers all over the world.}
    Each platform allows a user to upload an image of a clothing item and receive relevant recommendations for items that they can either buy on the same platform (like Amazon) or are presented as out-links to the corresponding seller's website (\textsc{Beagle.Vision} and \textsc{Lykdat}). All the three platforms detect multiple clothing items in the uploaded image and show the recommendations for each such item, with both \textsc{Beagle.Vision} and \textsc{LykDat} also displaying an associated label for each detected item. This improves explainability as it allows a user to interpret the platform's understanding about the clothing item in question. On the other hand, since Amazon does not share any labels, a user has no way of knowing the reason for irrelevant recommendations. 
	During our data collection, we discovered that Amazon did not identify dungarees worn by males; instead the service retrieved recommendations for overalls worn by shipyard workers and beekeepers. However, the service perfectly identified the dungarees worn by females. We point out here that these results are anecdotal and dependent on various factors like the IP address, the time and day of the search and the browser on which the search is performed and as such are subject to change with time and geographic location. One sample result for each platform on a non-binary suit as input is shown in Figure~\ref{img2: search-results}. As can be clearly seen, Amazon's recommendations are heavily male centric whereas both Beagle and Lykdat have female centric recommendations as well.

	\subsection{Demographics of the survey participants}
	\if{0}\begin{table}[t]
		\noindent
		\small
		\centering
		\begin{tabular}{ |c|c|c|c|}
			\hline
			{\bf Age Group} & {\bf Males} & {\bf Females} & {\bf Professional Background}\\
			\hline
			19-25 & 5 & 6 & Undergraduate College Student \\
			\hline
			26-30 & 1 & 2 & Research Scholar \& Working Professional\\
			\hline
			31-35 & 1 & 0 & Working Professional \\
			\hline
		\end{tabular}	
		\caption{\footnotesize {\bf Demographics of the participants who took part in the survey. The majority were undergraduate college students and females.}}
		\label{Tab: user-demographic}
		\vspace{-5 mm}
	\end{table}\fi

	We advertised a call for volunteers on various social media sites like LinkedIn and Twitter, and internal channels and were able to recruit 15 people who consented to participate in the study. Eight of them self-identified themselves as female and the other seven self-identified themselves as male. The participants belong to various demographics ranging from undergraduate college students to working professionals between the ages 19-35. One of the participants was classified as an \textit{expert} as the person had a background in fashion design whereas the other participants were classified as \textit{normal users} who could be potential buyers of clothing items on an e-commerce platform. A more detailed distribution of the participant demographics is noted in Table~\ref{Tab: user-demographic}.

	\subsection{Data collection and participant surveys}
	We divided our study into two parts -- \textit{data collection}, performed between July and September, 2021 and \textit{participant surveys}, conducted in September, 2021. 
	
	\noindent\textit{\textbf{Data collection}}: We automated our data collection procedure by using the selenium tool to upload the input images one by one, and collect all the recommendation responses for each input, on the Google Chrome browser from the same geographic location and the same IP address. For each input image, we collect the photo and the link to the product page displayed on the result page, in the order it is displayed. We carried out this procedure for all the three platforms. Note that these platforms are constantly changing their designs as well as (possibly) their recommendations which might slightly affect the results reported here.
	
	\noindent\textbf{\textit{Participant surveys}}: We created Google forms for our survey questions that were then shared with the participants to fill. Each form took $\sim 2$ hours to fill and had the following components.
	\begin{compactitem}
		\item Questions on personal details like name, email, age group, and consent. None of these were made available to anyone except the authors.
		\item 28 unique input images randomly chosen from our collected dataset, ensuring an equal number of images from each sub-category mentioned in Table~\ref{Tab: dataset-clothing}.
		\item For each input image, images of the \textit{top 5} products in the recommendation result (henceforth referred to as \textit{output image}), for each platform leading to a total of 15 output images arranged in a random order. If any item was not identified or there was no relevant recommendation due to misclassification, we did not include any results for that particular platform, thus leading lesser output images (5/10 images) for that input clothing item.
		\item For each output image, there were two questions in the form -- the first was an MCQ asking the participant to choose according to their personal judgement which gender the output image clothing most conformed to -- male only, female only or non-binary. The second question asked the participant to rate the recommended image on its relevance using a 5-point Likert like scale with rating points from 1 (least relevant) to 5 (most relevant). The exact questions were as follows (Figure~\ref{img2:d}).
		\begin{compactitem}
			\item \textit{Q1.} Which of the following gender options is the most relevant for the given piece of clothing?
			\item \textit{Q2.} Please choose the relevance score for your chosen recommendation.
		\end{compactitem}
	\end{compactitem}	
    	\section{Results and Discussion}
	We aggregate the responses to the two questions per output image by each participant and analyse the results for 112 input images and 1625 total output images. We evaluate the participant responses across the dimensions of the clothing categories created earlier -- male-centric, female-centric and non-binary across all platforms and also per online platform -- Amazon Stylesnap, Beagle.Vision and Lykdat.

	\subsection{Gender choice for clothing recommendations}
	
	\begin{table}[t]
		\noindent
		\small
		\centering
		\begin{tabular}{ |c|c|c|c|c|}
			\hline
			{\bf } & {\bf All } & {\bf Stylesnap} & {\bf Beagle.Vision} & {\bf Lykdat}\\
			\hline
			Male-centric & Male only (49.92$\%$) & Male only (56.22$\%$) & Non-binary (48.04$\%$) & Male only (50.90$\%$) \\
			\hline
			Female-centric & Female only (63.03$\%$) & Female only (68.42$\%$) & Female only (56.17$\%$) & Female only (64.14$\%$) \\
			\hline
			Non-binary & Non-binary (47.42$\%$) & Non-binary (49.15$\%$) & Non-binary (45.34$\%$) & Non-binary (47.81$\%$)\\
			\hline
		\end{tabular}	
		\caption{\footnotesize {\bf Most frequent choice of gender option for recommended clothing items aggregated over all input images and all participant responses. The rows correspond to the gender option chosen by us for the input clothing items and the individual cells correspond to the most frequent value chosen by the participants as well as their percentages.}}
		\label{Tab: gender-freq}
		\vspace{-5 mm}
	\end{table}
	
	We first evaluate the participant answers for the gender they feel is most relevant for the recommended image given an input image. The participant can choose one of three options -- male only, female only or non-binary. In Table~\ref{Tab: gender-freq}, we present the results for all inputs distributed as per Table~\ref{Tab: dataset-clothing}. Let us denote the set of images in our dataset by $D$. The results are calculated in the following manner -- we first take the modal value from all participant responses for each recommended image, thus getting a set of gender values $G_i$ for each input image $i \in D$. Then we take the modal values from each set $G_i$ and get the most frequent gender value response for each input image $i \in D$. Finally, we consider all inputs distributed in the three categories and calculate the most frequent gender choice per category. The same procedure is followed for the modal values from each platform. The rows in the table correspond to the gender option chosen by us for the input clothing items. The individual cells correspond to the most frequent value chosen by the participants along with their percentages.
	
	\noindent\textit{\textbf{Observations}}: 
	\blu{From the second column (titled ``All''),}
	we can see that roughly $50\%$ responses correspond to the gender category of the input image. This confirms our own assumptions about the category divisions of the input dataset. 
	In the columns corresponding to each platform we see that the percentage values are even more enhanced for each category of clothing with more people being confident about the gender choice for female-centric clothing as compared to male-centric or non-binary clothing. Interestingly, for Beagle.Vision we see that a majority of the responses point toward the clothing being non-binary instead of male-centric.
	
	\noindent\textit{\textbf{Takeaways}}: The following conclusions can be drawn from these results.
	\begin{compactitem}
		\item Amazon has the highest percentage of responses for the corresponding most frequent gender options compared to Beagle.Vision and Lykdat.
		\item The participants choose the non-binary option for the male-centric clothing recommendations from Beagle.Vision most often. Thus this platform may appeal more to people who are looking for non-binary clothing options.
		\item The responses for Amazon Stylesnap and Lykdat are quite similar.
	\end{compactitem}

	\subsection{Average rating for the recommended output}
	
	\begin{table}[t]
		\noindent
		\small
		\centering
		\begin{tabular}{ |c|c|c|c|c|}
			\hline
			{\bf } & {\bf All } & {\bf Stylesnap} & {\bf Beagle.Vision} & {\bf Lykdat}\\
			\hline
			Male-centric & \textbf{3.89} | 0.34 & \textbf{3.65} | 0.83 & 3.54 | 1.06 & \textbf{4.09} | 0.41 \\
			\hline
			Female-centric & 3.74 | 0.41 & 3.43 | 0.63 & \textbf{3.64} | 0.77 & 4.06 | 0.44 \\
			\hline
			Non-binary & 3.35 | 0.50 & 2.75 | 1.08 & 3.33 | 0.96 & 3.39 | 1.19 \\
			\hline
		\end{tabular}	
		\caption{\footnotesize{\bf Mean relevance rating and the standard deviation for recommended clothing items aggregated over all input images and all participant responses. The rows correspond to the gender option chosen by us for the input clothing items and the individual cells correspond to the mean ($\mu$) and standard deviation ($\sigma$) per dataset category averaged over input images. The highest value for each column are in bold.}}
		\label{Tab: gender-rating}
		\vspace{-5 mm}
	\end{table}
	
	We next evaluate the average ratings given by the participants to each recommended item image on its relevance to the input image. The participants rate each recommended image on a 5-point Likert like scale ranging from 1 (least relevant) to 5 (most relevant). The results for this experiment are presented in Table~\ref{Tab: gender-rating} and calculated as follows- We first take the average rating from all participant responses for each recommended image, thus getting a set of average ratings for each recommended image. Next we average these values across all the recommended images to obtain an average per input image $i \in D$. Finally we calculate the mean ($\mu$) and standard deviation ($\sigma$) for each dataset category to obtain the final values.
	
	\noindent\textit{\textbf{Observations}}: From the table we observe that the male-centric clothing items are the most relevant overall with the non-binary clothing items being the least relevant. This confirms our assumption that online shopping platforms, while being adept at recognizing and recommending relevant clothing items for binary genders, are still lagging behind when it comes to recommendations for non-binary clothing items. These differences are even more pronounced on the Amazon platform, with non-binary clothing items receiving a recommendation of only 2.75 as compared to 3.65 for male-centric clothing, a gap of almost 1 rating point. The ratings on Beagle.Vision are very close for the male and female-centric clothing items with non-binary items being off by 0.31 rating points from the female-centric items. Finally, we see that Lykdat has the highest average ratings amongst all the three platforms for all input categories, with male-centric clothing receiving an average rating as high as 4.09 out of 5. Although the average rating for non-binary clothing is highest amongst all platforms at 3.39, it is still 0.70 points lower than the binary gender items.
	
	\noindent\textit{\textbf{Takeaways}}: The following conclusions can be drawn from these results.
	\begin{compactitem}
		\item The average ratings for male-centric clothing are the highest and for non-binary clothing are the least.
		\item Although the non-binary gender choice is most frequent for Amazon (Table~\ref{Tab: gender-freq}), it has the lowest ratings for non-binary clothing amongst all categories and platforms. This means that although it shows clothes that may be more non-binary but they are not very relevant to the input being shown.
		\item Similarly, Lykdat has the highest average ratings for all categories, with the values for male-centric clothing being the highest amongst all categories and platforms, even though it has a lower percentage of responses choosing the male only option for its recommendations as compared to Amazon. 
	\end{compactitem}
	
	\subsection{Ground truth ranking order vis-a-vis participant ranking order}
	\begin{table}[t]
		\begin{minipage}{0.42\textwidth}
		\noindent
		\footnotesize
		\centering
		\begin{tabular}{ |c|c|c|c|}
			\hline
			{\bf } & {\bf Stylesnap} & {\bf Beagle.Vision} & {\bf Lykdat}\\
			\hline
			Male-centric & \textbf{0.102} & 0.080 & -0.038 \\
			\hline
			Female-centric & 0.007 & \textbf{0.106} & -0.115 \\
			\hline
			Non-binary & -0.003 & -0.040 & \textbf{0.114} \\
			\hline
		\end{tabular}	
		\caption{\footnotesize {\bf Average Spearman's rank correlation between the ground truth rank ordering and the participant rank ordering. The rows correspond to the gender option chosen by us for the input clothing items and the individual cells correspond to average Spearman's rank for the corresponding platform. The highest positive values for each column are in bold.}}
		\label{Tab: spearman-correlation}
		\vspace{-5 mm}
		\end{minipage}
		\begin{minipage}{0.42\textwidth}
		\noindent
	\footnotesize
	\centering
	\begin{tabular}{ |c|c|c|c|}
		\hline
		{\bf } & {\bf Stylesnap} & {\bf Beagle.Vision} & {\bf Lykdat}\\
		\hline
		Male-centric & \textbf{0.091} & 0.058 & -0.033 \\
		\hline
		Female-centric & -0.004 & \textbf{0.088} & -0.095 \\
		\hline
		Non-binary & -0.019 & -0.044 & \textbf{0.106}\\
		\hline
	\end{tabular}	
	\caption{\footnotesize {\bf Average Kendall's $\tau$-b rank correlation between the ground truth rank ordering and the participant rank ordering. The rows correspond to the gender option chosen by us for the input clothing items and the individual cells correspond to average Kendall's $\tau$-b rank for the corresponding platform. The highest positive values for each column are in bold.}}
	\label{Tab: kt-correlation}
	\vspace{-5 mm}
		\end{minipage}
	\end{table}

\if{0}	\begin{table}[t]
	\noindent
	\small
	\centering
	\begin{tabular}{ |c|c|c|c|}
		\hline
		{\bf } & {\bf Stylesnap} & {\bf Beagle.Vision} & {\bf Lykdat}\\
		\hline
		Male-centric & \textbf{0.091} & 0.058 & -0.033 \\
		\hline
		Female-centric & -0.004 & \textbf{0.088} & -0.095 \\
		\hline
		Non-binary & -0.019 & -0.044 & \textbf{0.106}\\
		\hline
	\end{tabular}	
	\caption{{\bf Average Kendall's $\tau$-b rank correlation between the ground truth rank ordering and the participant rank ordering. The rows correspond to the gender option chosen by us for the input clothing items and the individual cells correspond to average Kendall's $\tau$-b rank for the corresponding platform. The highest positive values for each column are in bold.}}
	\label{Tab: kt-correlation}
	\vspace{-5 mm}
	\end{table}\fi
	
	\begin{table}[t]
	\begin{minipage}{0.42\textwidth}
	\noindent
	\footnotesize
	\centering
	\begin{tabular}{|c|c|c|c|}
		\hline
		{\bf } & {\bf Stylesnap} & {\bf Beagle.Vision} & {\bf Lykdat}\\
		\hline
		Male-centric & 0.602 & \textbf{0.676} & 0.550 \\
		\hline
		Female-centric & \textbf{0.513} & 0.419 & 0.431 \\
		\hline
		Non-binary & 0.292 & 0.259 & \textbf{0.618} \\
		\hline
	\end{tabular}	
	\caption{\footnotesize{\bf Average Spearman's rank correlation between the expert's rank ordering and the normal participant rank ordering. The rows correspond to the gender option chosen by us for the input clothing items and the individual cells correspond to average Spearman's rank for the corresponding platform.  The highest positive values are in bold.}}
	\label{Tab: expert-correlation}
	\vspace{-5 mm}
	\end{minipage}
	\begin{minipage}{0.42\textwidth}
	\noindent
		\footnotesize
		\centering
		\begin{tabular}{ |c|c|c|}
			\hline
			{\bf } & {\bf M-centric item} & {\bf F-centric item}\\
			\hline
			\textbf{M participant} & M only (75.91$\%$) & F only (86.37$\%$) \\
			\hline
			\textbf{F participant} & M only (49.46$\%$) & F only (68.74$\%$) \\
			\hline
		\end{tabular}	
		\caption{\footnotesize{\bf Confusion matrix of participant's gender and their response for the gendered clothing items. The values in the cell describe the choice and the percentage of times the particular gender choice is chosen by the participants of that particular gender. Males have a more binary outlook in their choice of gender for the clothing recommendations compared to females. M: male, F: female.}}
		\label{Tab: participant-comp}
		\vspace{-5 mm}
	\end{minipage}

	\end{table}

	We use the participant ratings to calculate the ranks for the recommended outputs. The participants are shown the output images in a random order and asked to rate each image on its relevance. This rating is used to calculate a new ranking for each product, on a per platform basis. A higher average rating means a lower (better) rank and vice versa. We then calculate the Spearman's correlation and Kendall's $\tau$-b correlation by comparing the original ground truth ranking against the participant ranking order. 
	
	We also find the correlation between the rank ordering by the expert participant and a set of normal participants for a set of 28 input images and their corresponding 405 output recommendations. This is done to understand how the perspective of an expert differs from a normal user on a given visual search recommendation result.
	
	The results for the Spearman's correlation are shown in Table~\ref{Tab: spearman-correlation}, and the results for Kendall's $\tau$-b correlation are in Table~\ref{Tab: kt-correlation}. Finally, Table~\ref{Tab: expert-correlation} presents the correlation coefficients for the comparison between the expert and the normal user.
	
	\noindent\textit{\textbf{Observations}}: From Table~\ref{Tab: spearman-correlation} and Table~\ref{Tab: kt-correlation} we can see that the correlations follow a very distinct pattern. Amazon's Stylesnap platform has the highest positive correlation for male-centric clothing whereas it has negative correlations for both female-centric and non-binary clothing items (Table~\ref{Tab: kt-correlation}). Similarly, Beagle.Vision has the highest positive correlation for female-centric clothing items and negative correlation for non-binary items, and a lower positive correlation for the male-centric ones. Finally, Lykdat has the highest positive correlation for non-binary clothing items, and negative correlations for both male and female-centric clothing items. This is in stark contrast with the results in Table~\ref{Tab: gender-rating} where Lykdat reported the highest average ratings for all input clothing categories. More specifically, even though the average ratings for male and female-centric clothing were $> 4$, the correlations are negative. This implies that 
    \textit{\blu{even though the recommendations shown by Lykdat are most relevant, their ranking order is very poor and completely opposite to what the users would expect.}}
	
	In Table~\ref{Tab: expert-correlation}, we compare the ranking done by the expert participant with the rankings done by three other normal participants \blu{chosen randomly} and calculate the Spearman's correlation coefficient for the same. 
	\blu{Due to resource limitations, we were unable to survey a higher number of participants.}
	The average correlations here are higher than that observed for the overall set of images. Surprisingly, we see that Amazon has the highest positive correlation for female-centric items whereas Beagle.Vision has the highest positive correlation for male-centric items. Non-binary items have the highest positive correlation on Lykdat, as before.
	
	\noindent\textit{\textbf{Takeaways}}: We can make the following conclusion from the results discussed above.
	\begin{compactitem}
		\item Although Lykdat has the highest average ratings for all types of clothing items, it also has least correlation for both male and female-centric clothing items.
		\item Beagle.Vision has the most negative correlation for non-binary clothing items, despite having a higher average rating than Amazon Stylesnap for the same.
		\item The comparison between expert and normal participant shows that the correlation for Lykdat on non-binary items is very strong, but the results for Amazon and Beagle.Vision are interchanged for male and female-centric items respectively.
	\end{compactitem}
	
	\subsection{Female vs male participants}
	
\if{0}	\begin{table}[t]
		\noindent
		\small
		\centering
		\begin{tabular}{ |c|c|c|}
			\hline
			{\bf } & {\bf Male-centric item } & {\bf Female-centric item}\\
			\hline
			\textbf{Male participant} & M only (75.91$\%$) & F only (86.37$\%$) \\
			\hline
			\textbf{Female participant} & M only (49.46$\%$) & F only (68.74$\%$) \\
			\hline
		\end{tabular}	
		\caption{{\bf Confusion matrix of participant's gender and their response for the gendered clothing items. The values in the cell describe the choice and the percentage of times the particular gender choice is chosen by the participants of that particular gender. Males have a more binary outlook in their choice of gender for the clothing recommendations compared to females.}}
		\label{Tab: participant-comp}
		\vspace{-5 mm}
	\end{table}\fi

	In this section, we compare the responses from the male and female participants. The participants choose one of the three gender options for each recommended item's image. For the results noted in Table~\ref{Tab: participant-comp}, we perform the following steps to get the results - for each output image, we first calculate the modal response from both male and female participants. We then count the number of times each gender group selects `\textit{male only}' and `\textit{female only}' option across all input images per clothing category. Finally, to calculate the percentage values, we take a ratio of the count calculated previously with the total number of output images within that category.

	\noindent\textit{\textbf{Observations}}: From the table we observe that male participants choose the `male only' option $\sim 76\%$ of the times for the male-centric clothing compared to female participants who do it only 40$\%$ of the times. Similarly, male participants also choose the `female only' option $\sim 86\%$ of the times whereas female participants do it only 59$\%$ of the times.
	
	On further investigation, we also observe that female participants choose the `non-binary' option for male-centric clothing $\sim 76\%$ of the times, against only $\sim 37\%$ of the times by male participants. Similarly, for female-centric clothing, female participants choose the `non-binary' option $\sim 75\%$ of the times, as opposed to $\sim 26\%$ of the times by male participants. These differences are very significant and show a stark contrast in the behaviour of the male participants compared to the females. 
	
	\noindent\textit{\textbf{Takeaways}} The following conclusions can be drawn from these results.
	\begin{compactitem}
		\item Males tend to choose the binary gendered option -- `male only' and/or `female only' more frequently than females.
		\item Females choose the `non-binary' option more frequently for both male-centric and female-centric clothing as compared to the corresponding gender option, thus indicating that females interpret clothing as non-binary more often. 
		\item \blu{This is an example of user interaction bias~\cite{mehrabi2021survey} as the observations can be attributed to the self-selected biased behaviour from the users.} 
	\end{compactitem}

    \section{Overall insights pointing to deprivation of the non-binary users}	
	
	In this case study, we have evaluated three different e-commerce platforms -- Amazon, Beagle.Vision and Lykdat for their visual search recommendations and evaluated these recommendations through human surveys. With growing public debate surrounding LGBTQ+ rights, there has been a rising concern about creating an inclusive shopping experience for LGBTQ+ shoppers~\cite{youthki} and various fashion retailers have adopted or pivoted themselves to address this concern~\cite{fibre2fashion,cnbc}. While this has worked for a few e-commerce platforms~\cite{vue_ai}, large online marketplaces like Amazon have refrained from adopting strategies as proposed by~\cite{vue_ai,contentsquare}. 
	
	We have presented a comprehensive quantitative evaluation of the human surveys over recommendations for male-centric, female-centric and non-binary clothing items and made some important observations. Based on the results in Tables~\ref{Tab: gender-freq} and~\ref{Tab: gender-rating}, it is immediately clear that all platforms under question are more robust to male and female-centric clothing over non-binary ones. The absolute percentage values for non-binary options in Table~\ref{Tab: gender-freq} are always lower than the two binary genders which means that users are less inclined to pinpoint non-binary choices for the recommendations that they see.
	
	Delving deeper into the relevance of these recommendations, it is painfully clear from Table~\ref{Tab: gender-rating} that even if non-binary recommendations are shown to the users, these are not very relevant to the input, with Amazon's recommendations in particular receiving extremely low ratings. This brings to fore the issue of how inclusive or engaging an LGBTQ+ person's shopping experience would be on such a platform. Amazon being one of the world's largest e-commerce platforms, needs to take responsibility for this and attempt to set a benchmark that others may follow.
	
	Using the recommendation ratings as a proxy for the ranking of the products, we observe from Tables~\ref{Tab: spearman-correlation} and~\ref{Tab: kt-correlation} that only the Lykdat platform has positive correlation between the ground truth ranks and the participant ranks for non-binary clothing items. Thus, not just are Amazon's recommendations irrelevant, they are also returned in a haphazard manner, thereby, creating a rather unpleasant shopping experience for such shoppers. At the same time, Amazon has the highest correlation for male-centric clothing items, which indicates that the system as a whole isn't broken, but is just not designed to handle non-binary clothing inputs. Beagle.Vision has a similar performance as Amazon.
	
	Finally, on comparing the male and the female participant responses, we observe a very stark contrast in their perception of clothing as a whole. Table~\ref{Tab: participant-comp} shows that male participants have a more binary opinion about clothing items compared to females. Since buyer personas are one of the deciding factors for inclusive design~\cite{contentsquare}, e-commerce platforms have a lower incentive to serve to LGBTQ+ shoppers if the male participants are assumed to be representative of the general shopper. This in turn puts off LGBTQ+ shoppers from buying on these platforms and creates a cyclic system that is doomed to only increase the divide.
    \section{Final remarks}
		
	We hope that this case study will allow e-commerce platforms to take note of the growing changes in society and be more inclusive in how they identify clothing items and show recommendations. We would also like to see all e-commerce platforms involved in this study to incorporate a gender-neutral or non-binary clothing section in their catalogue so that it may serve as safe shopping space for people who may face discrimination and persecution in public for using physical shopping stores and/or changing rooms.
	
	In future we would like to incorporate other platforms and scale up the study in terms of more number of images and respondents. However, we believe that all our observations shall remain the same.

	\bibliographystyle{ACM-Reference-Format}
	\bibliography{main}
	
	
\end{document}